# Nanocellulose-stabilized Pickering emulsions: fabrication, stabilization, and food applications

**Chuye Ji, Yixiang Wang***

Department of Food Science and Agricultural Chemistry, McGill University, Ste Anne de Bellevue, Quebec, Canada H9X 3V9

*Correspondence: yixiang.wang@mcgill.ca, +001(514)3987922

**ABSTRACT**

Pickering emulsions have been widely studied due to their good stability and potential applications. Nanocellulose including cellulose nanocrystals (CNCs), cellulose nanofibrils (CNFs), and bacterial cellulose nanofibrils (BCNFs) has emerged as sustainable stabilizers/emulsifiers in food-related Pickering emulsions due to their favorable properties such as renewability, low toxicity, amphiphilicity, biocompatibility, and high aspect ratio. Nanocellulose can be widely obtained from different sources and extraction methods and can effectively stabilize Pickering emulsions via the irreversible adsorption onto oil-water interface. The synergistic effects of nanocellulose and other substances can further enhance the interfacial networks. The nanocellulose-based Pickering emulsions have potential food-related applications in delivery systems, food packaging materials, and fat substitutes. In this review, we highlight key fundamental work and recent reports on nanocellulose-based Pickering emulsion systems. The sources and extraction of nanocellulose and the fabrication of nanocellulose-based Pickering emulsions are briefly summarized. Furthermore, the synergistic stability and food-related applications of nanocellulose-stabilized Pickering emulsions are spotlighted.

## 1. Introduction

Pickering emulsions are solid particle-stabilized emulsions free of molecular surfactants, where solid particles are adsorbed onto the oil-water interface. Compared to the conventional emulsions, the irreversible adsorption of the nano-sized particles at the oil/water interface makes the Pickering emulsions preserve the droplets against coalescence and Ostwald ripening. In particular, there has been an increase in the replacement of common synthetic surfactants in food products with natural and sustainable substitutes such as plant-derived materials [1]. The demand for food-grade colloidal particles to formulate Pickering emulsions is growing because of the significant role of Pickering emulsions in food industries [2]. In multi-phase food Pickering emulsion systems, protein and polysaccharides particles can function as stabilizers in the form of nano-/micro-particles [3, 4]. However, many protein particles may cause structural instability and aggregation, and polysaccharide particles usually exhibit poor emulsifying performance and surface activity [5, 6].

The extraction, characterization, and application of nanocellulose (including CNCs, CNFs, and BCNFs) have been widely studied. According to the search results of Web of Science, 6648 research papers on nanocellulose have been published from 2017 to 2022, of which 175 papers (around 3%) are about Pickering emulsions. Nanocellulose possesses the unique properties of high aspect ratio, wettability, and good mechanical properties, which make it an excellent stabilizer of Pickering emulsions. Its emulsifying capacity can be further improved by physical/chemical modifications and/or the incorporation of other synergistic stabilizers, and the resultant nanocellulose-based Pickering emulsions have potential food-related applications such as the controlled release of bioactive compounds, construction of active food packaging materials, and substitution of fat in food. There is a lack of comprehensive overview of recent development in nanocellulose-based Pickering emulsions for food-related applications [7, 8]. Therefore, this review focuses on the recently reported nanocellulose-stabilized Pickering emulsions, and summarizes their raw materials, traditional/emerging technologies, and food-related applications. It aims to provide perspectives on future study to promote the full potential of nanocellulose-based Pickering emulsions.

## 2. Source and extraction of nanocellulose

Nanocellulose has been obtained from various sources, including woody and non-woody plants [9] (asparagus [10], bamboo pulp [11], cotton [11], eucalyptus [12], *Miscanthus floridulus* straw [13], and wood pulp [11]), marine organisms (tunicate [14] and algae [15]), and bacteria [16-19]. The extraction of CNCs and CNFs is usually based on top-down chemical and/or mechanical strategies, while BCNFs are directly synthesized by bacteria. Herein, the different sources and extraction methods of recently reported nanocellulose for the preparation of food-related Pickering emulsions are summarized in Table 1. Compared with woody biomass, agricultural resources contain a lower content of lignin, and thus require less chemicals and energy consumption during the extraction process [9]. Especially, agricultural wastes are excellent cellulosic feedstock because of their low-cost and sustainability, and much attention has been attracted to the extraction of nanocellulose from agricultural wastes, such as banana peels [20], defatted rice bran [21], ginkgo seed shells [22], lemon seeds [23, 24], oil palm fruit bunch [25], pistachio shells [26], and sweet potato residue [27]. Meanwhile, BCNFs can be more easily obtained from certain bacterial species (i.e., *Komagataeibacter xylinus* [16], *Gluconacetobacter xylinum* ATCC23767 [17], *Komagataeibacter hansenii* CGMCC 3917 [18], and *Acetobacter xylinum* CGMCC5173 [19]), because of the high purity of bacterial cellulose [28]. The extraction of CNCs usually consists of pretreatments (for removal of dirt and soluble substances), purifications (for removal of hemicellulose and lignin), and fragmentation (for generation of nanostructure via hydrolysis or mechanical destruction) [7, 29]. The use of recyclable chemicals, showing eco-friendly advantages, has been applied to extract nanocellulose, such as organic acid [30], solid acid [31], and recycled sulfuric acid [32]. However, strong acid hydrolysis causes the harsh equipment corrosion and environmental pollution and is less favorable. Generally, CNCs extracted by acid hydrolysis exhibit the higher crystallinity and shorter length than those by mechanical methods, and the properties of CNCs can be controlled and modulated by appropriate adjustments of hydrolysis or mechanical conditions [7]. Different from CNCs, CNFs can be prepared facilely through high-pressure homogenization, microfluidization [33], or ultrasonication. Particularly, CNFs prepared by aqueous counter collision (ACC-CNFs) showed enhanced exposure of inherent hydrophobic surface planes, which contributed to the better emulsifying capacity [34, 35]. The obtained nanocellulose can be further modified, for example by charged surfactants such as octenyl succinic anhydride (OSA) [36, 37], polymeric dialdehyde [38], cinnamoyl chloride, and butyryl chloride [39], to adjust its hydrophobicity [40].

## 3. Fabrication of nanocellulose-stabilized Pickering emulsions

Rotor-stator homogenization, high-pressure homogenization, and sonication are the most widely used techniques to prepare Pickering emulsions (Figure 1). Almost half of the Pickering emulsions presented in this review were prepared by the rotor-stator homogenizer (Tables 1 and 2), which consists of a rotor with blades and a stator with openings [41]. Rotor-stator homogenization has many advantages such as low operating cost, ease of setting-up, rapid process, and small amount of liquid required, but also drawbacks of the lack of uniformity, limited energy input, and risks of temperature increase and high shear rate [41-43]. High-pressure homogenizer has been frequently used in the industry and consists of a standard homogenizing nozzle and a high-pressure pump with pressure ranging from tens to hundreds of MPa. The primary coarse emulsion prepared at the pre-emulsification step is injected in the homogenizing nozzle to reduce the emulsion droplet size even down to the nanometer range. Compared with rotor-stator homogenization, high-pressure homogenization can prepare smaller droplets, process large volume samples in a continuous manner, and control droplet size by adjusting the pressure or the number of homogenizing cycles [41, 42]. However, the high-pressure homogenization has several disadvantages such as high operating cost, large minimum volume, difficulty of cleaning, increase of temperature, and high shear rate. High intensity ultrasound with frequencies greater than 20 kHz, as a green technology, has been utilized for emulsification [44]. The titanium probe transmits the ultrasonic energy to surrounding samples, inducing emulsification by cavitation and ultrasonic force. The ultrasonic emulsification possesses high energy efficiency, good emulsion stability, and ease of operating and cleaning [41, 45]. However, it suffers from the risks of the disruption of fragile particles or particle aggregates and the deposition of trace amount of titanium into the sample during ultrasonic emulsification. Besides these three techniques, membrane emulsification and microfluidization have also been applied to prepare Pickering emulsions, but the industrial scale production is still a challenge [46].

**Fig. 1** Schematic diagram of fabrication methods of nanocellulose-stabilized Pickering emulsions.

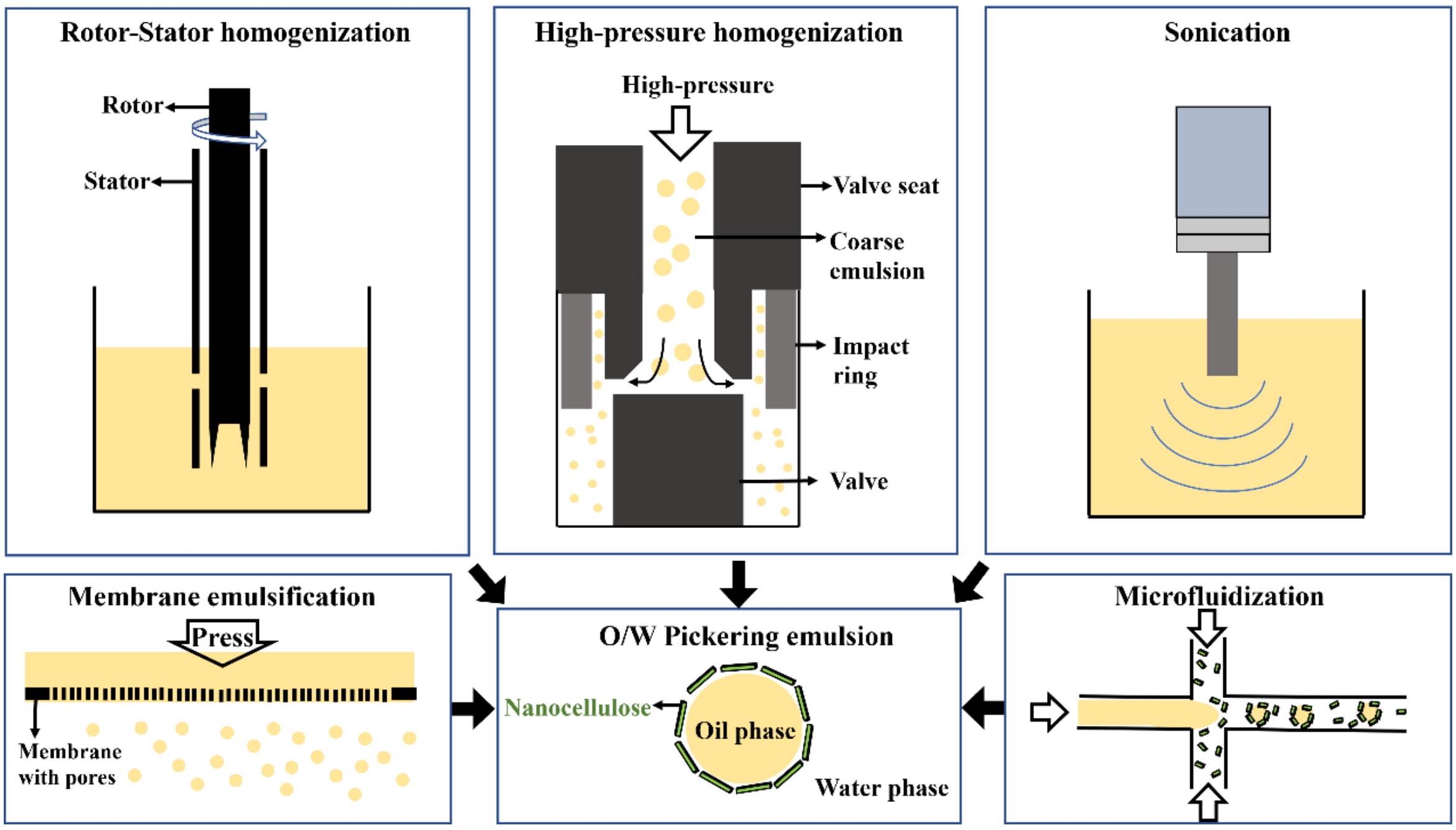

Pickering emulsions are stabilized by solid particles at the oil-water interface, showing higher resistance to droplet coalescence and better elastic responses than surfactant-stabilized emulsions [47]. These solid particles (stabilizers/emulsifiers) have partial wettability in both immiscible phases, and the energy required to remove a particle from the interface is orders of magnitude greater than the thermal energy [48]. The morphology of emulsifiers/stabilizers is important. For example, the non-spherical particles such as nanocellulose theoretically have stronger ability to generate a densely packed network at the interface because they have high energy of attachment even at a low concentration [49-51]. It has been proved that the longer particles with higher-aspect-ratio presented the larger emulsion volume and better stability [52]. Nanocellulose has been widely applied to stabilizing Pickering emulsions. CNCs can provide a steric barrier at the interface to prevent emulsion droplets from coalescence, while CNFs and BCNFs generate the network at the interface as well as in the continuous phase, resulting in the hindrance for coalescence and the increase of emulsion viscosity.

**Synergistic stability**

Nanocellulose is usually modified to improve the emulsifying capacity [53]. Compared to chemical modifications, the generation of cellulose-based complexes by hydrogen bonding and/or electrostatic interaction is an facile, cost-effective, and eco-friendly way to adjust the hydrophobicity and interfacial tension of nanocellulose [54]. The recently reported complexes of nanocellulose and other synergistic substances for food-related Pickering emulsions are summarized in Table 2. Liu et al. [18] reported the preparation of BCNF-soy protein isolate (SPI) composite colloidal particles via antisolvent approach. The addition of BCNFs exposed the hidden hydrophobic regions of SPI through hydrogen bonding and resulted in the improved viscoelasticity (gel-like behavior), crystallinity, thermal stability, and interfacial wettability. The high internal phase Pickering emulsions (HIPEs) stabilized by BCNF-SPI (7:25, w/w) composite particles exhibited high stability during 2-month storage. Similarly, the SPI-BCNF complexes were prepared by triggering the electrostatic interaction [54, 55]. The addition of SPI increased the surface hydrophobicity and emulsifying capacity of TEMPO-oxidized BCNFs, and allowed the formation of uniform and highly stable Pickering emulsions [55]. Moreover, the Pickering emulsions showed good oxidative stability because of the steric barrier built by SPI-BCNF complexes, and the increased viscosity of the emulsions inhibited the movement of oxidative free radicals. The SPI-TEMPO-oxidized BCNF complexes were also applied to construct edible foams with excellent mechanical properties, porous structure, and good energy absorption capacity [56]. Besides, CNC-bovine serum albumin (BSA) complexes were formed via strong electrostatic interaction at pH 3.0, which showed the increased stiffness, low surface charge, and improved amphiphilic properties, and were suitable for the formation of HIPEs [57].

Polysaccharides such as chitosan, chitin, and sodium alginate have also been employed as co-stabilizers of Pickering emulsions. Baek et al. [58] modified chitosan (GCh) with glycidyl trimethylammonium chloride (GTMAC) and formed the complexes with phosphorylated-CNC (P-CNC) by ionic gelation under mild sonication. The GCh-P-CNC nanocomplexes showed a hard sphere morphology and an average diameter of 300-500 nm, and the obtained Pickering emulsions were highly stable without creaming, coalescence, or phase separation during 3-month storage. A similar thick particulate layer formed by CNF-chitin complexes could also prevent the emulsion droplets from coalescence and creaming [59], while the complexes of high molecular weight sodium alginate (SA) and BCNCs had better amphipathicity compared to the original BCNCs [17]. Interestingly, besides the design of nanocellulose-based complexes, Zhou et al. [60] prepared the Pickering emulsions by blending anionic nanocellulose-stabilized droplets with cationic nanochitin-stabilized droplets. At high levels of nanochitin-coated droplets (≥50%), the cationic droplets formed a shell layer around the anionic droplets, resulting in the reduced viscosity and increased stability.

Lignin has functional properties such as anti-oxidation and ultraviolet adsorption, and has been applied to modulate the hydrophilicity of nanocellulose [61, 62]. Recently, Guo et al. [33] reported the preparation of CNFs containing residual lignin. The content of lignin on CNFs could be modulated during extraction process, and the emulsions with more residual lignin showed the higher emulsion ratio and stability. Polyphenols with promising bioactivities (antioxidative, antibacterial, anti-inflammatory, etc.) can bind to nanocellulose through hydrogen bonds and hydrophobic and cation-π interactions [63]. For example, the complexation of tea polyphenols increased the interfacial diffusion rate constant and surface activity of BCNFs, and the Pickering emulsions exhibited great free-radical scavenging activity [63], while the addition of tannic acid (TA) as a co-stabilizing agent condensed the interfacial layer and formed the insoluble shell with CNCs and methyl cellulose to realize the dispersibility of freeze-dried emulsions [64].

Besides them, food-grade surfactants can also promote the interfacial adsorption of nanocellulose [65]. Food-grade lauric arginate (LAE) was combined with CNCs at a mass ratio of 2:1 (LAE: CNCs), and the Pickering emulsions stabilized by positively charged LAE-CNCs complexes showed small droplet size (~1 μm) and good stability. Pelegrini et al. [66] created a novel green strategy to improve the hydrophobicity of CNCs by using triterpene saponins (SAP), which endowed the Pickering emulsions with improved mechanical resistance to coalescence and deformation.

Inorganic particles such as positively charged hectorite nanoparticles (AHNPs) could be combined with negatively charged BCNFs to effectively prevent the coalescence of emulsion droplets [67]. Meanwhile, the incorporation of $Fe_3O_4$ magnetic nanoparticles could not only improve the stability of Pickering emulsions [68], but also endow them with a dual responsive behavior towards magnetic field and pH changes [69]. Magnetic cellulose nanocrystals (MCNCs) were prepared by ultrasound-assisted *in-situ* coprecipitation and exhibited the increased wettability and higher surface charge compared to original CNCs [70, 71]. Mikhaylov et al. [72] found that the oil-in-water Pickering emulsions stabilized by $Fe_3O_4$-CNCs complexes possessed high dynamic viscosity and thixotropic properties, while the obtained palm olein-in-water Pickering emulsions were stable under pH 3-6 and maintained the morphology and size of emulsion droplets at pH 6 for at least 14 days [71]. The dense MCNC layer formed at oil/water interface acted as a physical steric barrier to preventing droplets from aggregation and coalescence, as well as showing a dual responsive behavior towards magnetic field and pH change.

**Double Pickering emulsions**

Double emulsions can be applied to improve the encapsulation and protection of multiple bioactive compounds with different solubilities [73]. They also provide the possibility of fabricating reduced-fat emulsion products by the replacement of conventional oil-in-water (o/w) emulsions with the equivalent water-in-oil-in-water (w/o/w) emulsions that have lower actual oil contents but similar in-mouth perceived texture [74]. Cunha et al. [75] designed novel surfactant-free oil-in-water-in-oil (o/w/o) Pickering emulsions stabilized by the combination of natural and partially esterified nanocellulose from softwood sulfite pulp fibers. The double emulsions were prepared by adding the modified nanocellulose suspensions in hexadecane to the primary o/w emulsions, exhibiting remarkable stability without coalescence during one-month storage [75].

## 4. Food-related applications

Due to their sustainability, biocompatibility, cost-efficiency, low toxicity, and high stability, nanocellulose-stabilized Pickering emulsions have many potential applications such as the delivery vehicles for bio-medicine design, pest control, and optoelectronic devices. Especially, more and more attention has been paid to their food-related applications, and recent research work has mainly focused on the encapsulation of bioactive compounds, construction of functional food packaging materials, and preparation of fat substitutes.

**Delivery systems of bioactive compounds**

The delivery systems with encapsulated bioactive compounds have promising applications in food and pharmaceutical fields, ranging from improving stability to controlling the release for enhanced functionality [76]. Qi et al. [77] prepared oleogel-in-water Pickering emulsions stabilized by CNCs to encapsulate β-carotene, which exhibited the remarkable stability over a wide range of pH (4-8) and salt concentrations (0.05-0.60 M). Patel et al. [78] developed Pickering emulsions from lauric acid modified CNCs for the encapsulation of β-carotene at different pH levels, dilution factors, and storage periods. The modified CNCs stabilized the morphology of lipid droplets and effectively prevented the creaming of lipid phase during storage. Wei et al. [79] designed the complexes of zein colloidal particles (ZCPs) and CNCs (the mass ratio of ZCPs and CNCs was 1:4) as the stabilizers for Pickering emulsions to encapsulate β-carotene (Fig. 2). A high retention rate (60.23%) was observed over 28-day storage period, and the multilayer structure could delay the lipolysis of emulsions during digestion and improve the bioaccessibility of β-carotene.

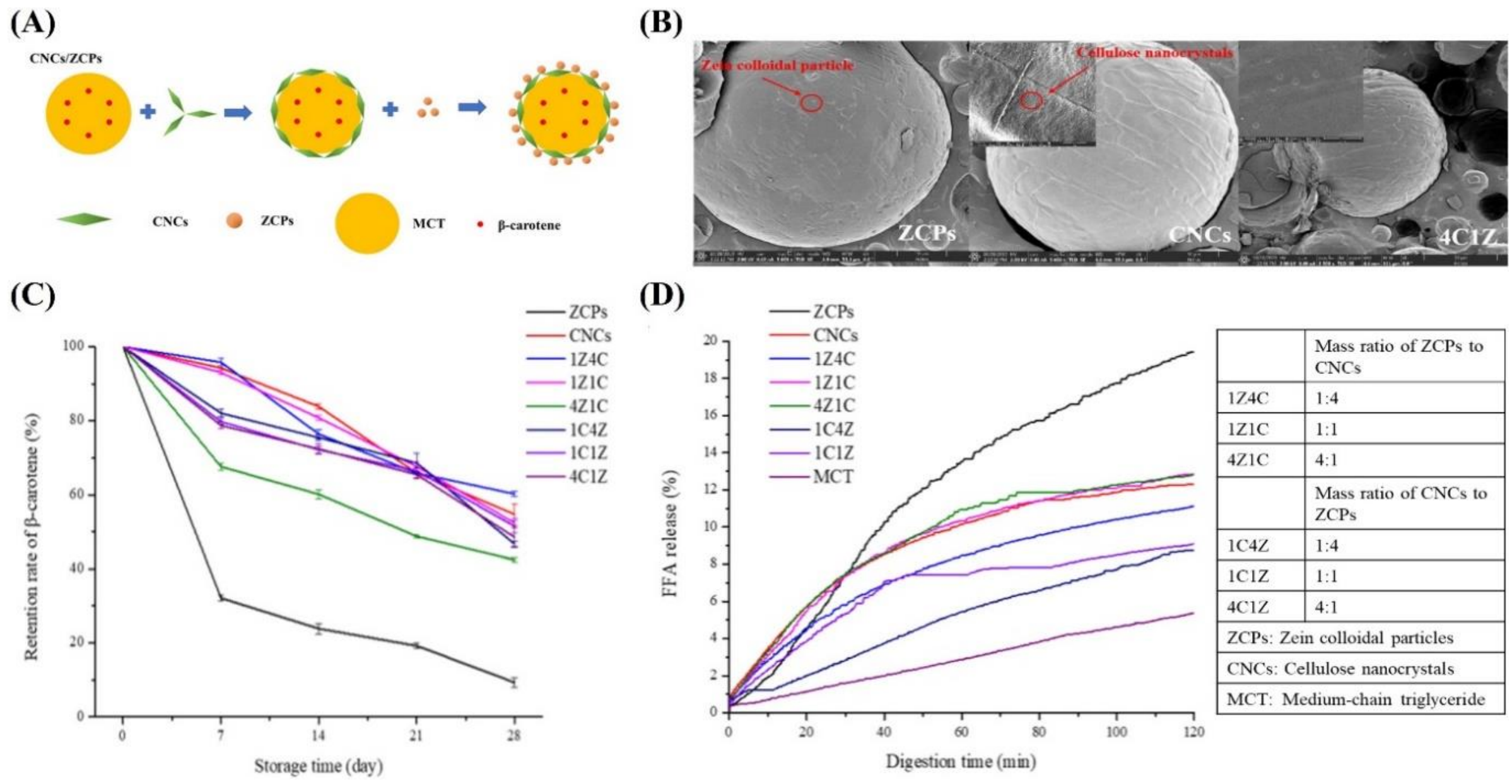

**Fig. 2** A) Schematic illustration of the fabrication of β-carotene loaded Pickering emulsion co-stabilized by ZCPs and CNCs; B) Cryo-SEM microstructures of different Pickering emulsions; C) Retention rate of β-carotene entrapped in different Pickering emulsions; D) Digestion time dependence of FFA release (%) from different Pickering emulsions. Modified from Wei et al. [79] with permission.

Curcumin has various biological and pharmacological activities (including antimicrobial, antioxidant, anti-inflammatory, and anticancer properties), but its therapeutic applications are limited because of its low stability and bioavailability. Recently, numerous research works reported that nanocellulose-stabilized Pickering emulsions could be used as delivery systems of curcumin [80-84]. Pickering emulsions stabilized by aminated CNCs for coumarin and curcumin showed high encapsulation efficiency (>90%), increased bioavailability, and low toxicity [80]. Some other Pickering emulsions stabilized by nanocellulose with encapsulated curcumin also exhibited good stability during long-term storage [83]. The stability of curcumin could be further improved with the incorporation of (−)-epigallocatechin-3-gallate (EGCG), which was added in the pineapple peel CNC-stabilized Pickering emulsions and interacted with CNCs through hydrogen bonding [85]. In the CNC/EGCG emulsions, curcumin was highly stable with 70% retention over a 30-day storage period and was not affected under thermal and ultraviolet light treatments. The release of encapsulated curcumin could be controlled by the incorporation of $Fe_3O_4$ [82]. Upon exposure to external magnetic field, the release rate of curcumin was accelerated and an inhibition of the growth of human colon cancer cells (down to 18%) was observed.

Vitamin $D_3$ is a required nutrient to control numerous important biochemical pathways but has poor water-solubility and low bioaccessibility. It could be encapsulated in CNF stabilized Pickering emulsions to improve the stability and bioaccessibility [86, 87]. The effect was related to the CNF concentration, where a significant decrease in the bioaccessibility and stability of vitamin $D_3$ was observed at the high level of CNF (7:10 NFC-to-oil) [86]. In the Pickering emulsions stabilized by CNC-pseudoboehmite heterocoagulates, 83% of initial vitamin $D_3$ was encapsulated and the target release in the small intestine was achieved [88].

Some other bioactive compounds were also employed as the models. For example, TEMPO-oxidized CNFs (TOCNFs) extracted form *Undaria pinnatifida*, one of brown algae, were used to stabilize astaxanthin (AXT)-loaded Pickering emulsions and enabled better storage stability and thermal stability (retention rate of AXT was 57.23% at 75°C) compared with the Pickering emulsions stabilized by whey protein isolate and xanthan gum (retention rate of AXT was about 30% at 70°C) [89]. BCNFs stabilized Pickering emulsions were applied to load and sustained release alfacalcidol (a hydrophobic drug) [19]. Lu et al. [90] used milled-cellulose-particles stabilized Pickering emulsions to deliver aged citrus peel extract, which exhibited a range of biological activities and improved bioaccessibility.

**Food packaging films**

Biopolymer-based edible films have gained much interest, but the hydrophilic nature of polysaccharides and proteins displays high moisture adsorption [91]. Pickering emulsions stabilized by nanocellulose have been applied as reinforcements or coatings to improve the functional properties of edible films, such as moisture barrier property, mechanical properties to resist the external destructive forces [92], antioxidant property to protect foods from oxidation [93], and antimicrobial property to prevent food spoilage [94-97]. Li et al. [98] prepared edible oleofilms by casting beeswax-in-water Pickering emulsions stabilized by hybrid particles of BCNFs and carboxymethyl chitosan (Fig. 3A). Beeswax as a natural wax is a remarkable barrier to moisture, while the addition of carboxymethyl chitosan enhanced the interactions within BCNFs network. With the increase of hybrid particle contents, the oleofilms showed the reduced water vapor permeability ($<1.1\times10^{-7}$ $g\cdot m^{-1}\cdot h^{-1}\cdot Pa^{-1}$), enhanced surface hydrophobicity (the contact angle increased from 113.1° to 139.1°), and improved re-dispersibility.

Deng et al. [99] prepared an edible coating from CNC-stabilized Pickering emulsions and chitosan (2 wt%). CNC Pickering emulsions enhanced the water vapor barrier property and stability of chitosan coatings, and an optimized oleic acid concentration of 20 g/kg retained peel chlorophyll, decreased senescent core browning, and delayed ethylene production and respiration peaks of postharvest pears [100]. Liu et al. [101] fabricated thermally curable Pickering emulsions stabilized by CNCs (3.0 wt%) (Fig. 3B). CNCs as a coating agent uniformly formed a film after drying and filled the pores of paper to realize a low permeability against air, contributing to enhanced tortuosity of water molecules through the film and raised water vapor barrier properties of the Pickering emulsion coating.

Moreover, nanocellulose was used as the reinforcement to provide structural support for nanocomposite polymer materials via the formation of nanocellulose-stabilized Pickering emulsions. Reinforcing pearl millet starch-based films with Kudzu CNCs increased tensile strength (from 3.9 to 16.7 MPa) and Young’s modulus (from 90 to 376 MPa) with the lower water vapor permeability (from 9.60 to 7.25 $\times10^{-10}$ $g\cdot m^{-1}\cdot s^{-1}\cdot Pa^{-1}$) [102]. Xu et al. [103] prepared dialdehyde-oxidized CNCs (DCNC) as stabilizers of Pickering emulsions. The DCNC-stabilized Pickering emulsions loaded with dihydromyricetin were incorporated into gelatin matrix to prepare gelatin-based edible films, resulting in strong UV-light barrier ability with high transparency, low water vapor permeability (~$3.62\times10^{-10}$ $g\cdot m^{-1}\cdot s^{-1}\cdot Pa^{-1}$), and enhanced mechanical property. Moreover, transparent films were obtained from gelatin and agar with the incorporation of clove essential oil Pickering emulsions stabilized by TEMPO-CNFs, and showed strong antioxidant activity and good UV blocking property without sacrificing the transparency ($T_{280}$ decreased from 26.9% to 1.7%) [96]. The polystyrene/CNF composite film prepared by encapsulation of a styrene monomer in CNF-stabilized Pickering emulsions showed good thermal dimensional stability and improved mechanical properties due to the reinforcement by homogenously distributed CNFs [104]. Liu et al. [105] developed high-barrier konjac glucomannan (KGM) based emulsion films by incorporating HIPEs, where HIPEs were stabilized by the complexes of BCNFs and soy protein isolation. With the increase of HIPEs concentration in KGM films, the HIPEs/KGM films exhibited the improved thermal stability, and the reduced moisture content, water solubility, water vapor permeability, and oxygen permeability.

Nanocellulose-stabilized Pickering emulsions are free of surfactants and have been considered as promising carriers for loading oil-soluble bioactive compounds to prepare edible films for keeping food quality and safety. Recently, researchers have successfully prepared novel antimicrobial packaging films by using nanocellulose-based solid particles, which contributed to the mechanical strength and stability of Pickering emulsions. Plant essential oils are natural bioactive compounds with excellent antimicrobial properties. The combination of cellulose particles and essential oils in Pickering emulsions provided a simple way to improve water vapor barrier property and antimicrobial activity. Wu et al. [106] fabricated CNF-based active films with remarkable antimicrobial and antioxidant activities by adding oregano essential oil Pickering emulsions stabilized by functionalized solid particles (Fig. 3C). Notably, the particles were complexes of polyethyleneimine (PEI) and TEMPO-oxidized CNCs (TOCNCs). The combination of PEI-TOCNCs (contact antimicrobial) and oregano essential oil (volatile antimicrobial) brought the enhanced antimicrobial properties (the inhibition rates: 97.28 ± 0.20% of *L.monocytogenes*, 97.23 ± 0.53% of *E.coli*) into CNF films. Souza et al. [94] fabricated starch-based thermoplastic films containing CNF-stabilized Pickering emulsions of ho wood essential oils, showing the improved mechanical resistance and less crystallinity (~2%) than the pristine starch films. The hydrogen bonding interactions between polyphenols of essential oils and amylopectin chains contributed to low melting temperatures (>50°C), showing a promising application in active food packaging to inhibit food spoilage and extend the shelf lives of foods.

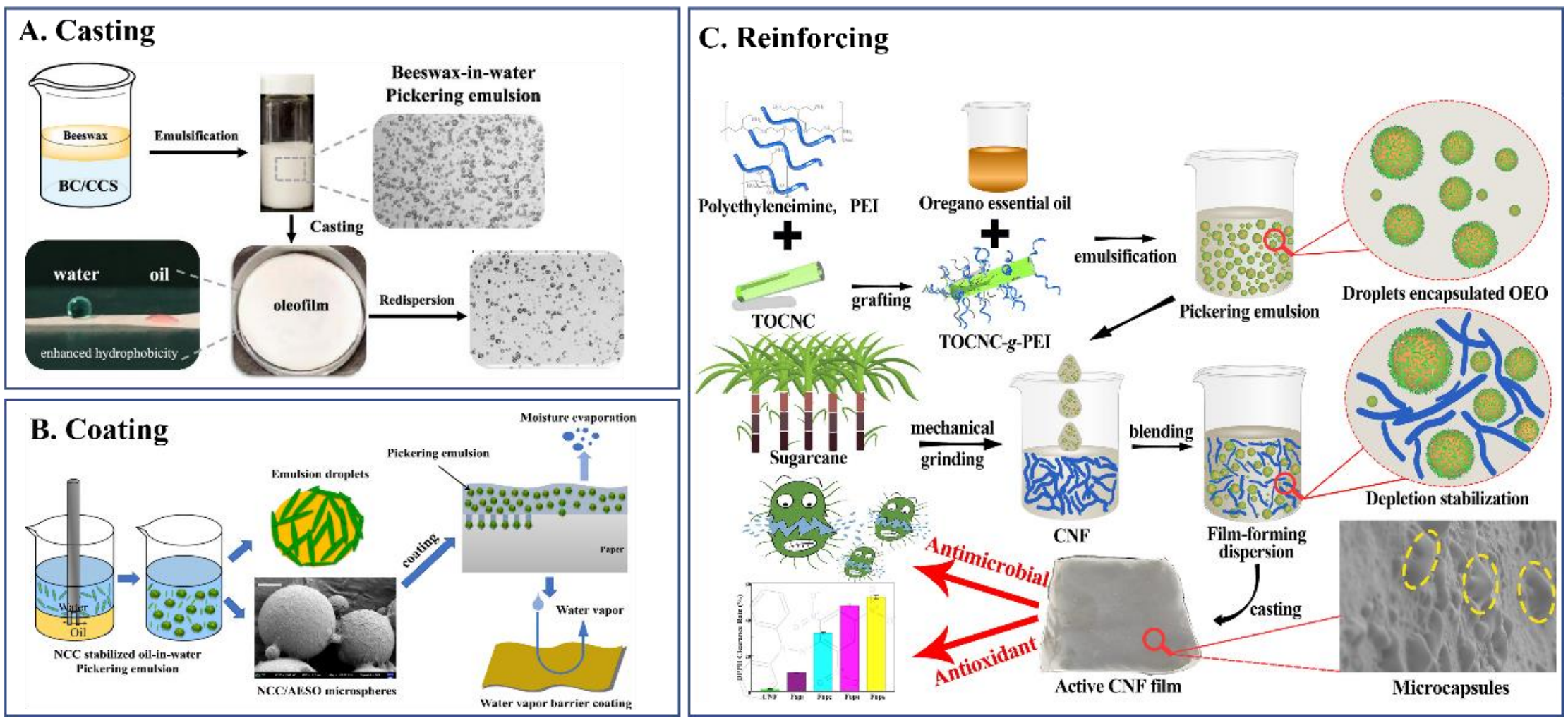

**Fig. 3** Schematic illustration of (A) the fabrication of beeswax-in-water Pickering emulsion and the casting process of oleofilm with enhanced hydrophobicity (reprinted from Li, et al. [98] with permission); (B) the preparation of CNCs-stabilized soybean oil-in-water Pickering emulsions and the emulsion coating process on paper (reprinted from Liu, et al. [101] with permission); (C) the preparation of antimicrobial and antioxidant CNF films containing microcapsules (reprinted from Wu et al. [106] with permission).

**Fat substitutes**

Fat substitutes can mimic the taste, functions, and sensory properties of fats and reduce the use of saturated and trans-fat in various foods. The utilization of fat substitutes to inhibit or slow down lipid digestion is an effective pathway to promote obesity reduction and decrease calorie intake. Recently, the application of Pickering emulsions as fat substitutes has attracted considerable interest [107]. Pickering emulsion-based foods are promising to decrease energy intake by altering the nutritional compositions, and nanocellulose has been considered as the structuring agent for liquid oil during emulsification to maintain the texture of food products. The physical barrier generated by the irreversible adsorption of nanocellulose at the oil/water interface can induce the resistance to pepsin and bile salt displacement [108, 109]. Moreover, nanocellulose can better prevent the diffusion of lipase to the droplet surface than the proteinaceous Pickering emulsion stabilizers, resulting in slow initial lipid digestion [108]. Therefore, nanocellulose-stabilized Pickering emulsions are expected to play an important role in the design of fat substitutes [110, 111]. It was worth noting that the lipid type influenced the digestion profiles of emulsions significantly. The rate and degree of lipolysis of Pickering emulsions with medium chain triglycerides (MCT) were greater than those with long chain triglycerides (such as soybean oil and canola oil) [83], and the restricted lipolysis could increase the bioavailability of lipid-soluble components because of the effective delivery to specific sites [110].

Recently, Le et al. [108] reported the Pickering emulsions stabilized by hydrophobically modified CNCs that decreased the release of short-chain fatty acids (SCFA) and showed the stability as a delivery vehicle for SCFAs to targeted regions of the digestive tract. CNCs are not responsive to human proteolytic enzymes, such as pepsin in the gastric phase. In addition, the resistance to bile salt displacement due to the electrostatic repulsion of CNCs to bile salts and the high desorption energy of irreversibly absorbed CNCs slowed the rate of initial lipid digestion. Zhang et al. [112] also confirmed that CNCs added to pea protein-particle-stabilized Pickering emulsions acted as a barrier and restricted the access of pepsin to the interfacial available substrate sites in the Pickering emulsions (Fig. 4A). Moreover, the composite layers of Pickering emulsions could modulate the intestinal digestion of oil droplets. Mackie et al. [113] found that CNCs were entrapped in the intestine mucosal layer (a barrier to nanoparticle uptake) and failed to reach the underlying epithelium, resulting in the reduction of lipid absorption. The elastic interface layer formed by BCNFs-SPI complexes also enhanced the anti-digestion of Pickering emulsions [55]. The adsorption of BCNFs generated physical hindrance to bile salts and lipase, resulting in reducing the possibility of lipolysis and delaying the lipid digestion. Liu et al. [114] found that TEMPO-oxidized CNFs-stabilized Pickering emulsions exhibited stable colloidal properties in the simulated intestinal environment and stronger inhibitory behavior to lipid digestion compared with Tween 80 and cholate. Meanwhile, the larger oil droplets reduced the surface area of triacylglycerol exposed to the lipase enzymes, which could contribute to the inhibition of lipid digestion [115].

Similarly, the coalescence and flocculation of droplets and the accumulation of free fatty acids at the droplet surfaces also resulted in the inhibition of lipolysis [115, 116]. Besides the above-mentioned factors, the length of cellulose nanoparticles could affect the *in vitro* gastrointestinal digestion of Pickering emulsions (Fig. 4B), where the Pickering emulsions stabilized by longer cellulose nanoparticles showed the lower release of free fatty acids due to the gel structure formed in the gastric stage [117]. A few potential applications have been evaluated. For example, Xie et al. [118] designed oil-in-water Pickering emulsions stabilized by bacterial cellulose and water-insoluble dietary fibers as fat alternatives to meet the demands of low-calorie baked foods (Fig. 4C). The reduction of oil content and the introduction of dietary fiber led to the formation of a fiber network and increased the viscoelasticity of the dough for chewy biscuit products. Similarly, Wang et al. [119] used

CNF-stabilized Pickering emulsions as fat alternatives in emulsified sausage. When 30% fat was replaced, the low-fat sausage exhibited brighter color and enhanced texture properties, such as higher hardness, springiness, and chewiness.

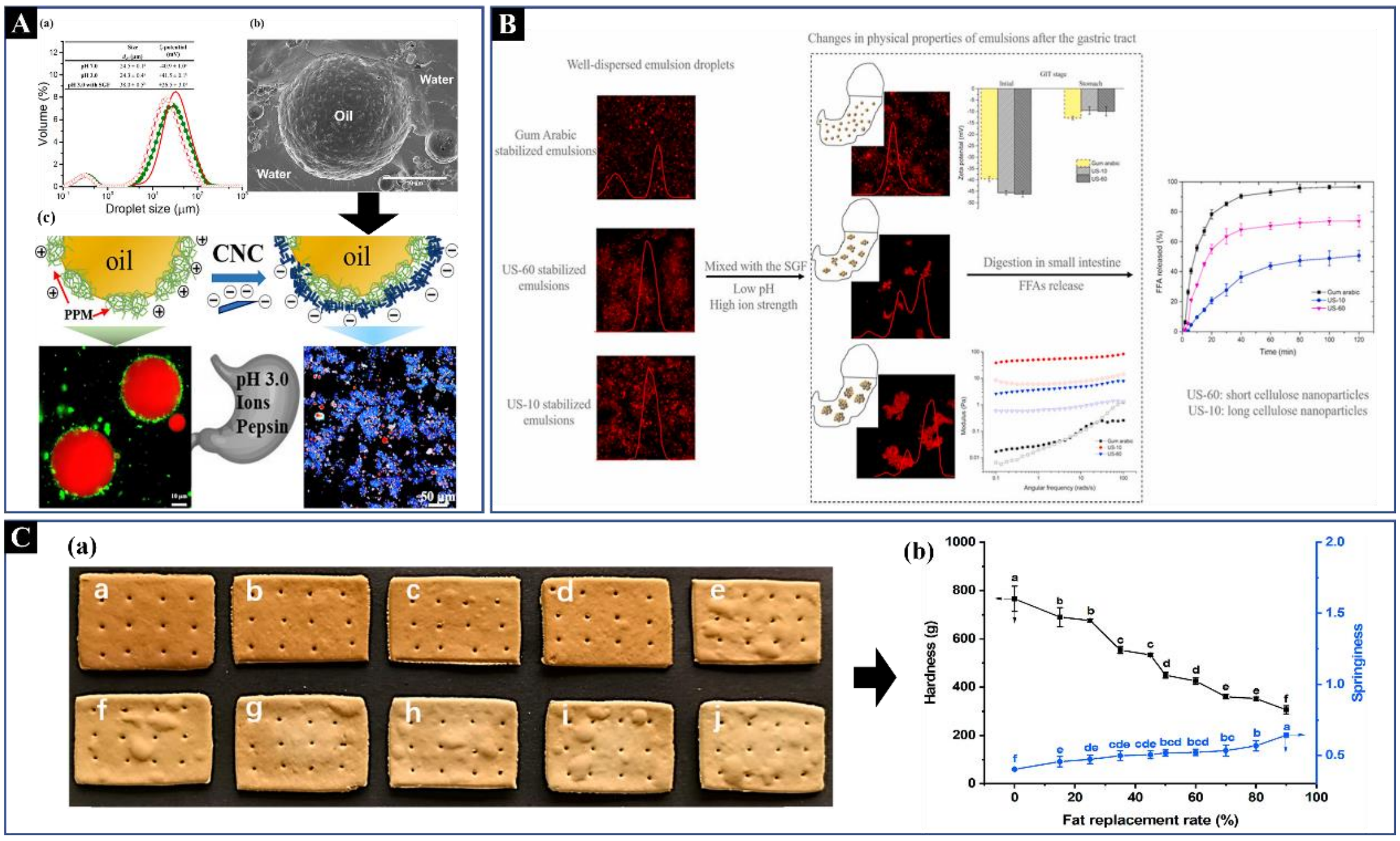

**Fig. 4** A: (a) Droplet size distribution of oil-in-water emulsions stabilized by PPM (PPM-E) at pH 7.0, pH 3.0 and PPM+SGF mixture at pH 3.0 without the addition of pepsin with insets showing corresponding volume-average mean diameter ($d_{43}$) and zeta-potential of all the samples. Different superscripts (a-c) in the same columns of the inset of (a) represent significant differences between different samples at $p<0.05$ level; (b) Cryo-SEM micrograph of PPM-E at pH 7.0. Scale bar in (b) represents 10 μm; (c) Schematic illustration of the gastric digestion of Pickering emulsions (reprinted from Zhang, et al. [112] with permission). B: Schematic illustration of the in vitro gastrointestinal digestion of nanocellulose-stabilized Pickering emulsions by altering cellulose lengths (reprinted from Ni et al. [117] with permission). C: (a) The color and appearance of biscuits prepared with different fat replacement rates; (b) The hardness and springiness of biscuit doughs prepared with different fat replacement rates. Different superscript letters in columns indicate statistically significant differences between groups ($p < 0.05$) (reprinted from Xie et al. [118] with permission).

## 5. Nanocellulose-stabilized Pickering emulsions for 3D printing

With the development of technology and materials, three-dimensional (3D) printing has revolutionized the conventional manufacturing. Pickering emulsions stabilized by nanocellulose have gained an increasing demand as suitable 3D printing inks due to their outstanding internal phase capacity and remarkable rheological behaviors (flowability as well as solidity against collapse) in terms of injectable and gelling properties [8, 120]. Nanocellulose stabilized HIPEs inks benefit from high storage modulus and shear-thinning property and are promising to print spatial architectures with high shape fidelity, high structural integrity, and low shrinkage [121, 122].

Ma et al. [123] prepared 0.5 wt% unmodified CNC-stabilized HIPEs with 80% internal phase fraction, which exhibited high printing resolution and fidelity at pH 7. The addition of NaCl could partially screen the surface charge of CNCs to promote their interfacial packing, and thus contribute to the stability of HIPEs [124]. The continuous phase of CNC-stabilized Pickering emulsions constructed by host-guest hydrogels enabled the high storage modules (reached up to ~113 kPa) and maintained the shear-thinning behavior, resulting in high-quality 3D printed structures with high shape fidelity, structural integrity, and excellent dimensional stability (volume shrinkage of 7 ± 2% after freeze-drying) [122]. Hydrocarbon chain grafted CNCs (HCNCs) stabilized HIPEs could be used for 3D printing to construct ultra-lightweight porous scaffolds with tunable structures by adjusting the oil phase fraction and aqueous conditions (such as pH value, ionic strength, and HCNCs content). Stimuli-responsive materials could be further incorporated in the printed hydrogels to control the release rate of internal oil phase in response to surrounding environment [125].

TEMPO-oxidized bacterial cellulose (TOBC) reinforced poly(lactic-acid) (PLA) was used to produce fully biodegradable nanocomposites through 3D printing [126]. The uniform dispersion of 1.5 wt% TOBC in PLA matrix improved the mechanical properties (tensile strength and elongation at break increased by 9.2% and 202%) and crystallization rate (increased by 116%) of PLA. Huan et al. [127] designed the two-phase emulgel systems for 3D printing that consisted of PLA/chitin/CNF Pickering emulsions and CNF/silica particles hydrogels (Fig. 5). The hierarchical layers generated by precise and simultaneous phase separation of immiscible but metastable emulsion/hydrogel phases allowed the regulation of diffusion and permeability of cargos, thereby controlling the mass transport and channeling to outer environment.

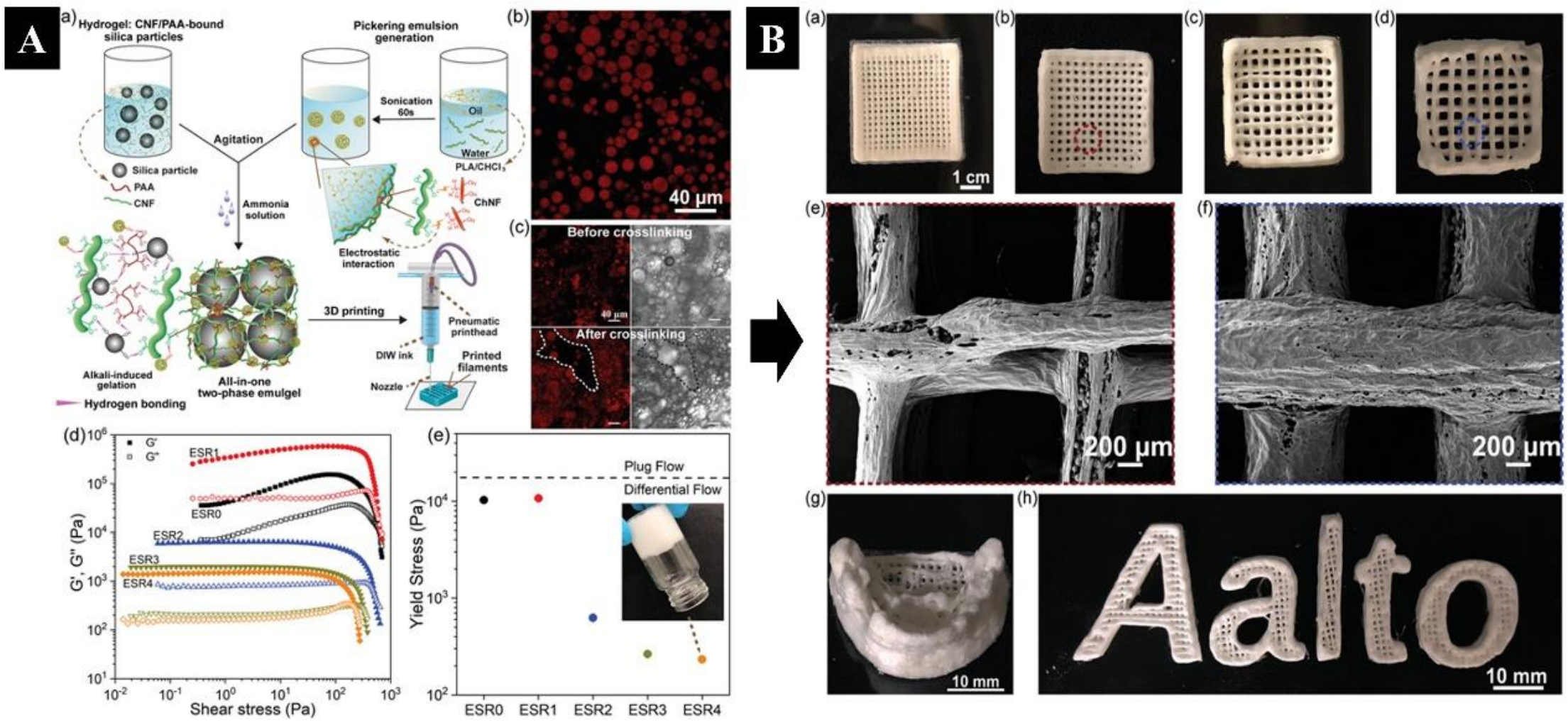

**Fig. 5** A: a) Schematic illustration (not to scale) of the fabrication of all-in-one, two-phase emulgels. b) Confocal image of PLA/CHCl3-in-water Pickering emulsion stabilized by nanofibrils isolated from cellulose (CNF) and chitin (ChNF), as shown in (a). c) Confocal and bright field images of two-phase emulgels before (upper) and after (bottom) alkali-induced gelation. The dashed line highlights a region depleted of droplets after gelation. d) Oscillatory rheology of the emulgels with increasing emulsion loadings. ESR0 indicates a system with no emulsion present. The emulsion-to-silica particle mass ratio in ESR1 to ESR4 correspond to 19:32, 29:22, 39:12, and 45:6, respectively. e) Shear yield stress of the emulgels, following the same color code used in (d). The added dashed line denotes the transition from differential to plug flow. The photo inset displays the ESR4 system in an inverted vial (2.5 cm diameter). B: Top view of cubic grids printed with ESR4 ink using a needle size of a) 0.25,

b) 0.41, c) 0.63, and d) 0.84 mm. SEM images of the grids printed by using e) 0.41 and f) 0.84 mm needle. Top view of the 3D-printed model of g) a human dental and h) the letters Aalto. Reprinted from Huan et al. [127] with permission.

## 6. Conclusions and perspective

Recently, Pickering emulsions stabilized by nanocellulose have shown increasing potential in various applications. The morphology and emulsifying capacity of nanocellulose are related to their original sources, pretreatment processes, and extraction methods, and the stability of Pickering emulsions could be further improved by synergistic stabilization. Nanocellulose-stabilized Pickering emulsions have offered promising advantages in food applications. Especially, they could improve the stability of encapsulated bioactive compounds and control the release for enhanced bioavailability. Functional food packaging films formed with nanocellulose-stabilized Pickering emulsions had good moisture barrier property, mechanical strength, antioxidant activities, and antimicrobial properties. Moreover, the physical barrier generated by the irreversible adsorption of nanocellulose at oil/water interface could slow down the digestion of lipid to promote obesity reduction. Nanocellulose also provided the excellent rheological properties of Pickering emulsions for 3D printing. Nanocellulose-stabilized HIPEs inks had high storage modulus and shear-thinning property, which enabled the printed architectures with high shape fidelity and structural integrity.

In future studies, the utilization of agricultural waste/by-products and the development of “green” and facile extraction methods for nanocellulose production deserve more attention. The current utmost challenge is to maintain the stability of nanocellulose-based Pickering emulsions in multi-component food systems and at various conditions (different pH values, temperatures, ionic strengths, etc.). Considering the controllable digestion of the loaded lipid, it is promising to develop composite interface layer based on rational design to further delay lipid digestion and/or site-dependently release lipidic bioactive molecules. It has been reported that soluble dietary fibers can promote metabolic benefits on body weight and glucose control [128], and higher intakes (25 ~ 29 g per day) of dietary fiber are suggested to protect against cardiovascular diseases, colorectal and breast cancer [129]. Despite the great promise, more studies on translating the knowledge of nanocellulose-based Pickering emulsions into new and useful applications are still expected. Additionally, it is worth noting that CNCs and BCNFs have not been listed as the "generally regarded as safe" substances by the food and drug administration (FDA) [130]. The case-by-case studies on the potential safety issues of nanocellulose-based Pickering emulsions need to be carried out.

## Declaration of competing interest

The authors declare that they have no conflicts of interest to this work.

## Acknowledgments

We would like to acknowledge financial support from the Natural Sciences and Engineering Research Council of Canada (NSERC RGPIN-2019-04498), and Natural Sciences and Engineering Research Council of Canada Discovery Launch Supplement (NSERC DGECR-2019-00472). C.J. would like to thank the China Scholarship Council (CSC NO. 202106790028) for financial support of her Ph.D. program.

Table 1. Summary of recent Pickering emulsions stabilized by nanocellulose for food-related applications. (N/R = not reported)

| Cellulose sources | | Extraction of nanocellulose | Size of nanocellulose | Oil phase [ratio of oil to water] | Emulsification methods | Refs. |
|---|---|---|---|---|---|---|
| Lignocellulosic sources (Woody and non-woody plants) | Asparagus | Sulfuric acid hydrolysis | Diameters of 178.2-261.8 nm | Palm oil [30:70] (v/v) | Sonication | [10] |
| | Bamboo pulp<br>Cotton fabric<br>Wood pulp | Phosphoric acid hydrolysis, high-pressure homogenization | Diameters of tens of nanometers | Caprylic/capric triglyceride [10:90-50:50] (v/w) | Sonication | [11] |
| | Eucalyptus fibers | High-pressure homogenization, high-intensity ultrasound | Diameter of 236 nm, length higher than 1 μm | Soybean oil [10:90] (v/v) | Sonication | [12] |
| | *Miscanthus floridulus* straw | High-pressure homogenization | Width of 33.27-49.79 nm, height of 3.68-7.55 nm | Dodecane [10:90] (v/v) | Homogenization | [13] |
| | Banana peels | High-pressure homogenization, high-intensity ultrasound | Diameters of 3.3-3.5 nm, length of 1.5-3.4 μm | Sunflower oil [10:90] (w/w) | High-pressure homogenization or sonication | [20] |
| | Defatted rice bran | Sulfuric acid (55 wt%) hydrolysis under sonication process | Diameter of 5.7-12.7 nm, length of 152-327 nm | Rice bran oil [50:50] (v/v) | Homogenization or sonication | [21] |
| | Ginkgo seed shells | Sulfuric acid (62 wt%) hydrolysis, high-pressure homogenization | Length of 406-1500 nm | Corn oil [10:90-70:30] (v/v) | High-pressure homogenization | [22] |
| | Lemon seeds | Sulfuric acid (64 wt%) hydrolysis | Diameter of 14.43 nm, length of 155 nm | Sunflower oil [50:50] (v/v) | Homogenization | [23] |
| | | Ball milling using ionic liquids [BMIM]Cl | Diameter of 42.84 nm, length of 1782 nm | | | |
| | Lemon seeds | Sulfuric acid hydrolysis (64 wt%), APS (1M) and TEMPO oxidation | Diameter of 26-42 nm, length of 340-380 nm | Sunflower oil [50:50] (w/w) | Homogenization | [24] |
| | Oil palm fruit bunch | Soda pulping, bleaching, TEMPO oxidation, and high-pressure homogenization | Width of 4 nm, length of few microns | Dodecane [20:80] (w/w) | Sonication | [25] |
| | Pistachio shells | Hydrolysis with 3 M HCl, Modification with OSA | Average diameter of 68.8 nm | Corn oil [10:90] (v/v) | Sonication | [26] |
| | Sweet potato residue | Dealing with deionized water, methanol, and DMAC | N/R | MCT [10:90] (v/v) | Homogenization | [27] |
| Algae | Macroalgae (*Eucheuma cottonii*) | Phosphoric acid hydrolysis | Diameters of 17-157 nm | Paraffin oil [10:90] (v/v) | Homogenization | [15] |
| Bacteria | *Komagataeibacter xylinus* | Disintegration using a blender at 15,000 rpm for 5 min, hydrolysis with 2.5 M HCl at 70°C for 1 h | Diameters of 30-80 nm, length of 100 nm-several μm | Peanut oil [15:85] (v/v) | High-pressure homogenization | [16] |
| | *Gluconacetobacter xylinum* ATCC 23767 | Activation of bacterial cellulose at 35°C for 7 days | Average fiber diameters of 56.2-74.1 nm | Olive oil [20:80] (v/v) | Sonication | [17] |
| | *Komagataeibacter hansenii* CGMCC 3917 | Disintegration using a blender at 15,000 rpm for 5 min, hydrolysis with 2.5 M HCl at 70°C for 1 h | Diameters of 947.70-1409.47 nm | Sunflower seed oil [75:25] (v/v) | Homogenization | [18] |
| | *Acetobacter xylinum* CGMCC 5173 | Sulfuric acid hydrolysis | Diameters of 259.6 nm | $CH_2Cl_2$ [10:90] (v/v) | Sonication | [19] |

Table 2. Summary of recent Pickering emulsions stabilized synergistically by nanocellulose and co-stabilizers.

| Co-stabilizers | | Nanocellulose | Oil phase [ratio of oil to water] | Droplet diameter | Emulsification methods | Refs. |
|---|---|---|---|---|---|---|
| Protein | Soy protein isolate | BCNFs | Sunflower seed oil [75:25] (v/v) | 947.70-1409.47 nm | Homogenization | [18] |
| | Soy protein isolate | BCNFs | Canola oil [50:50] (v/v) | 10-40 μm | High-pressure homogenization | [55] |
| | Soy protein isolate | BCNFs | Dodecane [20:80-50:50] (v/v) | 10-15 μm | Homogenization | [56] |
| | Soybean protein | CNCs | Soy oil [20:80-50:50] (v/v) | < 0.1 μm | Homogenization | [131] |
| | Soybean protein isolate | BCNFs | Dodecane [50:50-74:26] (v/v) | 10-15 μm at pH 3.0; 15-40 μm at pH 7.0 | High-pressure homogenization | [132] |
| | Soy protein isolate | Bacterial cellulose | Sunflower seed oil [74:26] (v/v) | 20.71±2.03 μm | Homogenization | [105] |
| | Bovine serum albumin | CNCs | Soy oil [80:20] (v/v) | 52-84 μm | Homogenization | [57] |
| | Peanut protein isolate | CNCs | Rapeseed oil [30:70] (v/v) | 5-7 μm | High-pressure homogenization | [133] |
| Polysaccharide | Chitosan | CNCs | Oleic acid [3:97] (v/v) | 888±127 nm | Homogenization | [99] |
| | GTMAC-Chitosan | Phosphorylated CNCs | Olive oil [30:70] (v/v) | 3-5 μm | Sonication | [58] |
| | Chitin | CNFs | Corn oil [10:90] (v/v) | < 45 μm | Sonication | [60] |
| | Chitin nanofibril | CNFs | Corn oil [10:90] (v/v) | 3.1-7.6 μm | Sonication | [59] |
| | Sodium alginate | BCNCs | Olive oil [20:80] (v/v) | 2.6-4.1 μm | Sonication | [17] |
| Inorganic particle | $Fe_3O_4$ | CNCs | Flaxseed oil [10:90] (v/v) | 1.14-2.68 μm | Homogenization | [70] |
| | $Fe_3O_4$ | CNCs | Paraffin oil [30:70] (v/v) | 2.9-3.3 μm | Sonication | [72] |
| | $Fe_3O_4$ | CNCs | Palm oil [30:70] (v/v) | 11.90-109.00 μm | Sonication | [71] |
| | Hectorite nanoplatelets | BCNFs | Silicone oil [10:90-90:10] (v/v) | 2~5 μm | High-pressure homogenization | [67] |
| Polyphenol | Tea polyphenols | BCNFs | Dodecane (or camellia seed oil) [10:90] (v/v) | < 30 μm | Homogenization | [63] |
| | Tannic acid | CNCs and methyl cellulose | Corn oil [20:80] (v/v) | < 5 μm | Sonication | [64] |
| | (−)-Epigallocatechin-3-gallate (EGCG) | CNCs | Corn oil [40:60] (v/v) | 7.9-9.45 μm | Sonication | [85] |
| Lignin | | CNFs | Dodecane [20:80] (v/v) | ~16 μm | Sonication | [33] |
| | | CNCs | Dichloromethane [16.7:83.3] (v/v) | < 30 μm | Homogenization | [61] |
| Surfactant | Lauric arginate (food grade) | CNCs | Soybean oil [5:95] (v/v) | ~1 μm | Homogenization | [65] |
| | Triterpene saponins (green) | CNCs | Soapberry oil [30:70] (v/v) | 221.2-581.3 nm | Sonication | [66] |